\documentclass{iopart}

\usepackage{graphicx}           % for eps figures
\usepackage{amssymb}            % for extra symbols

\begin{document}

\letter{Zener quantum dot spin filter in a carbon nanotube}

\author{D~Gunlycke$^1$, J~H~Jefferson$^2$, S~W~D~Bailey$^3$, C~J~Lambert$^3$, D~G~Pettifor$^1$ and G~A~D~Briggs$^1$}

\address{$^1$ Department of Materials, University of Oxford, Oxford, OX1~3PH, UK}

\address{$^2$ QinetiQ, St. Andrews Road, Malvern, WR14~3PS, UK}

\address{$^3$ Department of Physics, University of Lancaster, Lancaster, LA1 4YB, UK}

\begin{abstract}
We predict and analyse a novel spin filter in semiconducting carbon nanotubes. By using local electrostatic gates, the conduction and valence bands can be modulated to form a double-barrier structure. The confined region below the valence band defines a Zener quantum dot, which exhibits resonant tunnelling. The resonances split in a magnetic field to make a bipolar spin filter for applications in spintronics and quantum information processing. We model this using $\vec{k}\cdot\vec{p}$ envelope function theory and show that this is in excellent agreement with a corresponding tight-binding calculation. 
\end{abstract}


\submitto{\JPCM}

A spin filter based on electrostatic gating of single-wall carbon nanotubes (SWCNTs) may have significant advantages over alternative approaches, based on spin injection from ferromagnetic contacts \cite{Tsukagoshi99,Lindelof02}.  Experimental techniques are now well established for placing source drain contacts at each end of a carbon nanotube and using a side gate to control electron transport.  Such devices exhibit Coulomb blockade in metallic \cite{Tans97,Tans98} and semiconducting \cite{Jarillo04} SWCNTs, and field-effect transistor action \cite{Tans98+,Javey03} at room temperature.  Electrical conductance through near ideal Ohmic contacts has been predicted \cite{White98} and shown experimentally to exhibit ballistic conductance close to the clean quantum limit of $4e^2/h$ \cite{Liang01,Kong01,Javey03}.  Electron interference was first demonstrated in metallic SWCNTs \cite{Liang01,Kong01} and more recently within the conduction band of semiconducting SWCNTs \cite{Jarillo04}, where the existence of Schottky barriers posed additional difficulties with contact resistance \cite{Javey03}.  The Kondo effect has been exhibited in metallic SWCNTs \cite{Nygard00} and optical emission from ambipolar FETs has been demonstrated \cite{Misewich03}.  All these devices have used global gating along the entire length of the active part of the nanotube.  Local gating of carbon nanotubes has been demonstrated with split gates, finger gates and top gates fabricated by lithographic means \cite{Robinson03,Wind03,Biercuk04,Mason04}.  Through local gating, the nanotube can be made to function as a pn-junction diode without modulation doping by using two closely spaced finger gates, which are positively and negatively charged with respect to source-drain contacts \cite{Lee04}.  Top gates have been used to form and control a double quantum dot \cite{Mason04}.

\begin{figure}
  \begin{center}
    \includegraphics[scale=0.75]{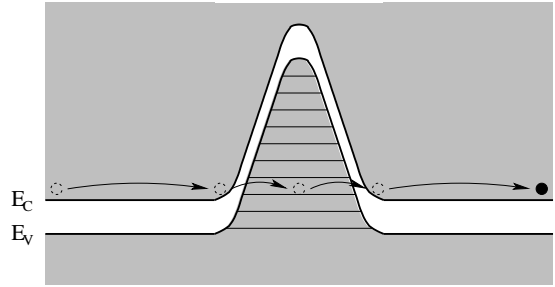}
  \end{center}
  \caption{Schematic diagram of the Zener resonant tunnelling.  In the gated region there are quasi-bound states below the valance band.  An incident electron with the same energy as one of these levels will be resonantly transmitted through.}
  \label{fig:1}
\end{figure}

\begin{figure}
  \begin{center}
    \includegraphics[scale=0.75]{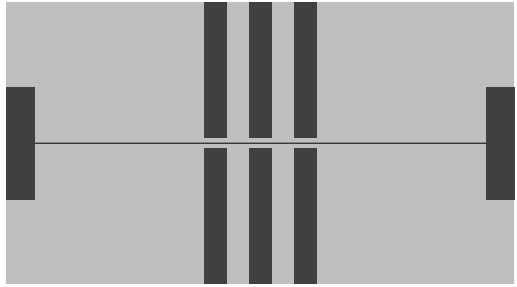}
  \end{center}
  \caption{Schematic diagram of the setup of gates and electrodes.  A pair of source and drain contacts are attached to the nanotube (center line).  Three pairs of split gates are situated in the top and bottom part of the figure.  Under the surface there is also a global backgate which cannot be seen in the diagram.}
  \label{fig:2}
\end{figure}

In this Letter we demonstrate that local gating also opens up the possibility of spin filtering via Zener tunnelling.  Although Zener tunnelling has yet to be demonstrated in semiconducting SWCNTs, our calculations show that there is no fundamental obstacle to its observation.  The main challenge is to achieve sufficiently localised electric fields along the nanotubes, through which an electron can tunnel elastically between valence and conduction bands.  If a locally-gated semiconducting SWCNT is gated negatively with respect to adjacent electrodes, the electrostatic field bends the energy bands upwards in the region of the gate, leading to a double Zener tunnelling structure and the formation of quasi-bound states, as illustrated in \fref{fig:1}.
In equilibrium these are occupied by holes above the Fermi level at low temperatures.  If an electron in the conduction band has energy corresponding to one of the quasi-bound states, resonant tunnelling will occur.  The resulting structure is essentially a quantum dot (QD), since there is confinement in all three spatial dimensions.  We emphasise that our Zener QD, inspired by previous work on a single-electron transistor \cite{Patent}, is an original idea which may have several applications in addition to our suggested spin filter.  Unlike a conventional dot, both the conduction and valence bands form the two confinement barriers using a single gate electrode.  These barriers provide an opportunity to aviod any material interfaces which generally inhibit electron coherence effects.  The gate electrode can be realised by using a local split gate \cite{Robinson03}.  For practical implementation we consider three split gates, as shown in \fref{fig:2}, in which the outer two are initially held at the same potential as the global back gate.
Their purpose is to confine the electrostatic potential, enabling higher electric fields along the nanotube.  Split gates of this kind have already been fabricated for multiwalled nanotubes \cite{Robinson03}.  These gates can be produced down to a width and lateral separation of order $20$ nm, so that a few volts applied to the electrodes give local fields of up to $2$ MV cm$^{-1}$ to enable Zener tunnelling.

In a perfect sheet of graphene there are two inequivalent Fermi points in the first BZ with wavevectors $\vec{K},\vec{K'}$.  These Fermi points are uncoupled in our system since the applied potential varies slowly on an atomic scale, and we can therefore apply $\vec{k}\cdot\vec{p}$ envelope function theory to the two Fermi points independently.  To derive an effective Hamiltonian we first expand the single-particle $\pi$-orbital Hamiltonian around an arbitrary Fermi point of graphene \cite{Slonczewski58}.  The dispersion for wavevectors close to this Fermi point is then given by the effective Hamiltonian:
\begin{equation}
  \label{eq:kp1}
  H^{(0)} = -i\hbar v_F \vec{\sigma}\cdot\vec{\partial},
\end{equation}
where $\{\sigma_i\}_{i=x,y}$ are the Pauli matrices, and $-i\vec{\partial} = -i(\partial_x,\partial_y)^T$ is the electron quasimomentum operator along and around the nanotube.  The pseudo-spin $\vec{\sigma}$ connects the A and B sublattices and accounts for both conduction and valence sub-bands with the real spin giving rise to further degeneracy which we do not consider explicitly yet.  This Hamiltonian is identical to Weyl's Hamiltonian with two spatial dimensions and solution of the corresponding Schr\"odinger equation with plane waves yields directly the energy dispersion, 
\begin{equation}
  \label{eq:dispersion}
  \epsilon(k_x,k_y) = \pm\hbar v_F\sqrt{k_x^2+k_y^2},
  \end{equation}
which  is the familiar light-cone in which the speed of light has been replaced by the Fermi velocity of graphene, $v_F \approx (1/300)c$.  In SWCNTs $k_y$ is quantised.  For metallic nanotubes $k_y = 0$ is allowed and equation \eref{eq:dispersion} yields the characteristic linear dispersion, $\epsilon(k_x) = \hbar v_F k_x$ of massless particles.  In a semiconducting zigzag nanotube the lowest quantised $k_y$ is finite and may be positive or negative, where the sign reflects the choice of Fermi point.  From equation \eref{eq:dispersion}  with $k_x=0$, we see directly that $k_y = \pm E_g/2\hbar v_F$, and hence the dispersion in the $x$-direction has the relativistic form of particles with rest mass (effective mass) $E_g/2v_F^2$. We only consider the lowest quantised bands since we will later show that Zener tunnelling through higher bands is negligible due to larger band separations.  These energy dispersions and the effective Hamiltonian \eref{eq:kp1} incorporate the rapidly varying atomic potentials due to the carbon atoms. Envelope function theory allows us to simply add to this the slowly varying applied potential, $V(x)$, to yield an effective Hamiltonian for the whole system, which is valid for the conduction and valence bands close to the Fermi energy. Since $V$ is independent of $y$ we may integrate out the $y$-motion for the lowest transverse channels, yielding the one-dimensional effective Hamiltonians, 
\begin{equation}
  \label{eq:kptot}
  H_\pm = -i\hbar v_F\sigma_x \partial_x \pm\frac{E_g}{2}\sigma_y+eV(x)I,
\end{equation}
where the positive and negative signs correspond to clockwise and anticlockwise motion around the tube respectively, and $I$ is the unit matrix.  We need only solve the Schr\"odinger equation for one of these cases since the degenerate sets of solutions are related by the transformation $\psi_- = \sigma_x \psi_+$.  By solving Laplace's equation the potential $V(x)$ for the gate configuration shown in \fref{fig:2} can be represented by the convenient analytic form
\begin{equation}
  \label{eq:Potential}
  V(x) = V_0 \cos^2\left(\frac{\pi x}{4a}\right),
\end{equation}
to a good approximation.  For the simulations described below, we have used the potential parameters $V_0=4$ V and $a=20$ nm.  We have assumed that there is no potential gradient transverse to the nanotube, which is a good approximation due to the symmetry of the split gates and the high ratio of gate width to tube diameter.

\begin{figure}
  \begin{center}
    \includegraphics[scale=0.5]{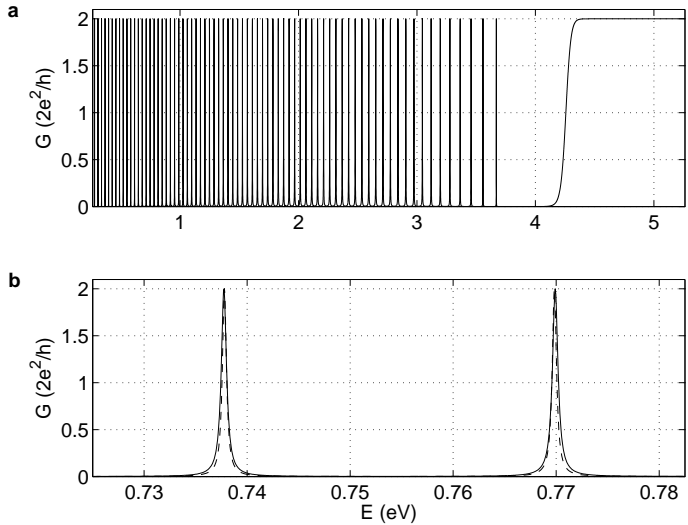}
  \end{center}
  \caption{Conductance as a function of single-electron energy in a (19,0)-nanotube.  (a) Small bias conductance derived from $\vec{k}\cdot\vec{p}$ envelope function theory.  The local gating has produced several sharp resonances.  (b) Small bias conductance at low energy comparing resonances from $\vec{k}\cdot\vec{p}$ envelope function theory calculations (dashed line) and Tight-Binding calculations (solid line).  Note that the Tight-Binding resonances confirm that only two channels contribute to the transmission.}
  \label{fig:3}
\end{figure}

We have solved the one-electron scattering problem using a nearest-neighbour finite-difference method and then calculate the low-bias conductance from the tunnelling transmission probability using the Landauer-B\"uttiker formula, $G=\frac{2e^2}{h}{\rm Tr}~tt^{\dagger}\equiv\frac{2e^2}{h}N(E)T(E)$, where $N(E)$ is the number of open channels and $T(E)$ is the average transmission probability per channel.  In \fref{fig:3}(a), we show results for the dimensionless conductance in the presence of the gate potential, obtained using the $\vec{k}\cdot\vec{p}$ envelope function approach.  This shows clear Zener resonances as a function of the electron energy $E$.  As a check on the validity of this long-wavelength approximation, we have computed the conductance using a tight-binding model using a recursive Green function approach \cite{Sanvito99} including all $\pi$-orbital bands from the quantisation of the graphene lattice.  The resulting Zener resonances from the two methods are compared in \fref{fig:3}(b) where we have zoomed the energy to show the separation and shape of two resonances at low energy.  The agreement is remarkable, with the very small deviations between the two methods arising from the difference in energy dispersion for high $k_x$-values on the dot.  In these calculations, four open channels are present ($N(E)=4$) for the chosen energy range, but only the lowest conduction and highest valence bands contribute to the resonant transport ($N(E)=2$), with almost total reflection from all other bands.

\begin{figure}
  \begin{center}
    \includegraphics[scale=0.5]{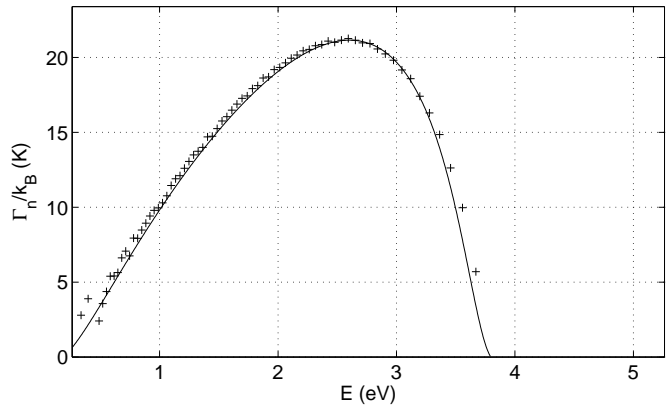}
  \end{center}
  \caption{The width of a resonance is plotted as a function of the single particle energy at the resonance.  The (+) markings represent the widths for the resonances calculated numerically using $\vec{k}\cdot\vec{p}$ envelope function theory.  The solid line is an estimate of the widths derived analytically using the WKB approximation.  In this figure the widths are calculated for a (19,0)-nanotube.}
  \label{fig:4}
\end{figure}

We can gain insight into these results by an approximate Breit-Wigner analysis of resonances through the two Zener barriers, which are described by the equation
\begin{equation}
  \label{eq:Breit_Wigner}
  T(E) = \frac{(\Gamma_n/2)^2}{(\Gamma_n/2)^2+(E-E_n)^2},
\end{equation}
where $E_n$ is the resonance energy of the $n$-th quasi-bound state and $\Gamma_n$ is the corresponding  half-width.  The latter can be estimated using a triangular potential \cite{TA} within the semiclassical WKB approximation \cite{WKB}.  For a nanotube with band gap $E_g$, the resonance width becomes
\begin{equation}
  \label{eq:WKB_resonance_width}
  \Gamma_n = \beta(E_n) \exp\left(-\frac{4E_g}{3\hbar v_F} w(E_n)\right),
\end{equation}
where the tunnel barrier width is $w(E_n) = L_+ - L_-$, with classical turning points $L_{\pm} \equiv \frac{4a}{\pi}\arccos\sqrt{\frac{E_n\mp E_g/2}{eV_0}}$.  The prefactor $\beta(E_n) \approx h\bar{v}_x(E_n)/L(E_n)$ mostly depends on the resonance energy via the confinement length $L(E_n) = 2L_-$.  $\bar{v}_x$ is the average group velocity along the nanotube direction.  In \fref{fig:4}, we have compared the resonance widths from equation \eref{eq:WKB_resonance_width} with the numerical widths derived from the $\vec{k}\cdot\vec{p}$ results [\textit{cf} \fref{fig:3}(a)].
In order to be resolved, the widths $\Gamma_n$ of the resonances must be larger than the thermal energy $k_BT$.  For our chosen potential [see equation \eref{eq:Potential}], the temperature of $(19,0)$-nanotubes must not be higher than of order 2-3 Kelvins at the conduction band edge.  The precise requirement is sensitive to the potential shape at low energies.  Generally, larger-diameter nanotubes with smaller band gaps exhibit larger resonance widths in accordance with equation \eref{eq:WKB_resonance_width}.  The band gap dependence is in fact the reason why Zener resonant tunnelling through higher subbands is strongly suppressed.

The location of the resonance energies $E_n$, may also be estimated using the WKB approximation, through
\begin{equation}
  \label{eq:resonance_condition}
  (n+\frac{1}{2})\pi = \int_{-L_-}^{L_-}k_n(x)dx.
\end{equation}
Apart from the turning points, where the wavevector is small and the integral is negligible, the energy dispersion is mainly in the linear regime, $\epsilon(k_x) \approx -\hbar v_F k_x$.  Writing the total energy as a sum of kinetic and potential energy, $E = \epsilon + eV$, where the potential energy is the applied potential in equation \eref{eq:Potential}, allows us to integrate equation \eref{eq:resonance_condition};
\begin{equation}
  \label{eq:res}
  \begin{array}{ll}
  \hbar v_F(n+\frac{1}{2})\pi = & \frac{4a}{\pi}\sqrt{\left(E_n+\frac{E_g}{2}\right)\left(eV_0-E_n-\frac{E_g}{2}\right)}\\
  & +\left(eV_0-2E_n-E_g\right) L_-.
  \end{array}
\end{equation}
If we Taylor expand the applied potential, $V(x)\approx V_0[1-(\pi x/4a)^2]$ and then integrate equation \eref{eq:resonance_condition}, we can solve for the resonance energies;
\begin{equation}
  \label{eq:res_Parabolic}
  E_n = eV_0-\frac{E_g}{2}-\left(\frac{3\hbar v_F\sqrt{eV_0}(n+\frac{1}{2})\pi^2}{16a}\right)^{2/3}.
\end{equation}
As a consequence of the linear dispersion, the resonance energies are $E_n \varpropto n^{2/3}$ instead of the usual $E_n \varpropto n$, for large $n$.  This phenomenon can be seen in \fref{fig:3}(a), where the resonances become more closely spaced towards the left of the diagrams, corresponding to increasing kinetic energies in the valence band.
\begin{figure}
  \begin{center}
    \includegraphics[scale=0.5]{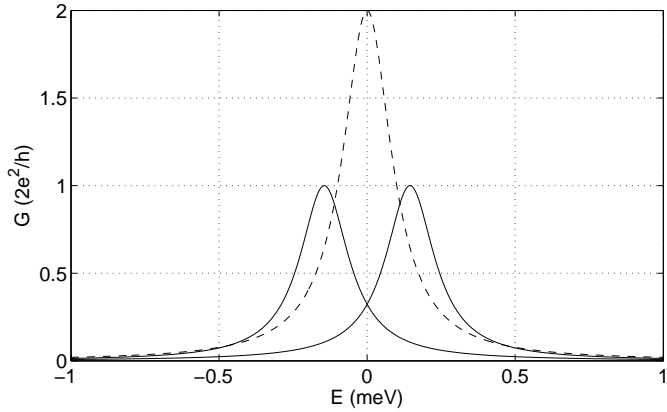}
  \end{center}
  \caption{Spin dependent resonances.  The dashed curve shows a resonance with the width $0.2$ meV.  The left(right) solid curve show the spin up(down) resonance arising when a transverse magnetic field of $5$ T is applied.}
  \label{fig:5}
\end{figure}
We now demonstrate that the above Zener tunnelling effect opens up the possibility of spin filtering for possible applications in quantum information processing, as has been suggested for unipolar quantum dots \cite{Recher00}.  In a uniform transverse magnetic field spin up/down electrons each experience their own set of resonances separated by $g\mu_BB \approx 0.29$ meV for $g \approx 2$ (\textit{cf} reference \cite{Jarillo04}) and $B=5$ T.  In \fref{fig:5} we have shown this for a resonance width of $0.2$ meV.  For these parameters, the polarisation is $P=(T_\uparrow-T_\downarrow)/(T_\uparrow+T_\downarrow) \approx 80\%$.  Higher polarisations can be achieved by using narrower resonances.  

This new spin filter has advantages over spin injection from ferromagnetic contacts, since spin-memory loss due to interface scattering at interfaces is avoided.  Local gating also allows complementary devices using electron- or hole-conduction, which may be advantageous if CN-based electronics is ever to form a basis for future nanoscale CMOS technology.

Finally we remark that Coulomb repulsion between the holes on the Zener QD will modify the resonant level structure.  This shifts the energy levels and, in particular, introduces an energy difference between electrons of opposite spin in the same orbital state.  In this sequential tunnelling regime the number of holes in the Zener dot will fluctuate between even and odd giving rise to a single-electron current from source to drain.  Consider a number of holes $n$ of the dot.  The hole levels will be occupied down to the resonant level.  An electron can resonantly Zener tunnel into the hole state (equivalent to the hole tunnelling from the dot to the source lead) leaving the dot with $n-1$ holes.  A further electron cannot tunnel into the dot since there will no longer be an available state to occupy.  Due to the resonant tunnelling the electron will then preferentially tunnel from the dot into the drain lead giving rise to a single-electron transfer from source to drain.

The presence of the Zener resonances could be demonstrated experimentally using a back-gate to change the Fermi energy which may be tuned to the resonances.  The source-drain voltage should be sufficiently small in order that the electrons do not exceed the optic phonon threshold of approximately $160$ meV.  In addition to semiconducting nanotubes, one may also attempt to use quasi-metallic nanotubes with much smaller bandgaps induced by curvature.  These have even smaller tunnel barriers, and consequently larger resonances which may be easier to observe.

\ack
This research is part of the QIP IRC (GR/S82176/01, is supported through the Foresight LINK Award Nanoelectronics at the Quantum Edge by EPSRC (GR/R660029/01, Hitachi Europe Ltd and EPSRC BT project 'Controlled electron transport through single molecules'.  This work has also been funded by the EU network MRTN-CT-2003-504574. GADB thanks EPSRC for a Professorial Research Fellowship (GR/S15808/01).  JHJ acknowledges support from the UK MOD.


\section*{References}

\begin{thebibliography}{12}

\bibitem{Tsukagoshi99}
Tsukagoshi~K, Alphenaar~B~W and Ago~H 1999 {\it Nature} {\bf 401} 572

\bibitem{Lindelof02}
Lindelof~P~E, Borggreen~J, Jensen~A, Nyg\aa rd~J and Poulsen~P~R 2002 Nobel Symposium {\it Physica Scripta} {\bf T102} 22

\bibitem{Tans97}
Tans~S~J, Devoret~M~H, Dai~H~J, Thess~A, Smalley~R~E, Geerligs~L~J and Dekker~C 1997 {\it Nature} {\bf 386} 474

\bibitem{Tans98}
Tans~S~J, Devoret~M~H, Groeneveld~R~J~A and Dekker~C 1998 {\it Nature} {\bf 394} 761

\bibitem{Jarillo04}
Jarillo-Herrero~P, Sapmaz~S, Dekker~C, Kouwenhoven~L~P and van der Zant~H~S~J 2004 {\it Nature} {\bf 429} 389

\bibitem{Tans98+}
Tans~S~J, Verschueren~A~R~M and Dekker~C 1998 {\it Nature} {\bf 393} 49

\bibitem{Javey03}
Javey~A, Guo~J, Wang~Q, Lundstrom~M and Dai~H~J 2003 {\it Nature} {\bf 424} 654

\bibitem{White98}
White~C~T and Todorov~T~N 1998 {\it Nature} {\bf 393} 240

\bibitem{Liang01}
Liang~W, Bockrath~M, Bozovic~D, Hafner~J~H, Tinkham~M and Park~H 2001 {\it Nature} {\bf 411} 665

\bibitem{Kong01}
Kong~J, Yenilmez~E, Tombler~T~W, Kim~W, Dai~H, Laughlin~R~B, Liu~L, Jayanthi~C~S and Wu~S~Y 2001 \PRL {\bf 87} 106801

\bibitem{Nygard00}
Nyg\aa rd~J, Cobden~D~H and Lindelof~P~E 2000 {\it Nature} {\bf 408} 342

\bibitem{Misewich03}
Misewich~J~A, Martel~R, Avouris~Ph, Tang~J~C, Heinze~S and Tersoff~J 2003 {\it Science} {\bf 300} 783

\bibitem{Robinson03}
Robinson~L~A~W, Lee~S~-B, Teo~K~B~K, Chhowalla~M, Amaratunga~G~A~J, Milne~W~I, Williams~D~A, Hasko~D~G and Ahmed~H 2003 {\it Nanotechnology} {\bf 14} 290; {\it Microelec. Eng.} {\bf 67} 615

\bibitem{Wind03}
Wind~S~J, Appenzeller~J and Avouris~Ph 2003 \PRL {\bf 91} 058301

\bibitem{Biercuk04}
Biercuk~M~J, Mason~N and Marcus~C~M 2004 {\it Nano Lett.} {\bf 4} 1

\bibitem{Mason04}
Mason~N, Biercuk~M~J and Marcus~C~M 2004 {\it Science} {\bf 303} 655

\bibitem{Lee04}
Lee~J~U, Gipp~P~P and Heller~C~M 2004 {\it Appl. Phys. Lett.} {\bf 85} 145

\bibitem{Patent}
Jefferson~J~H and Phillips~T~J 1998 DERA patent {\it The Zener Single-Electron Transistor}

\bibitem{Slonczewski58}
Slonczewski~J and Weiss~P 1958 \PR {\bf 109} 272

\bibitem{Sanvito99}
Sanvito~S, Lambert~C~J, Jefferson~J~H and Bratkovsky~A~M 1999 \PR B {\bf 59} 11936

\bibitem{TA}
Sze~J~M 1981 {\it Physics of Semiconductor Devices} (John Wiley: New York) 2nd Edition

\bibitem{WKB}
Wentzel~G, 1926 \ZP {\bf 38} 518; Kramers~H~A 1926 \ZP {\bf 39} 828; Brillouin~L 1926 {\it CR Acad. Sci.} {\bf 183} 24

\bibitem{Recher00}
Recher~P, Sukhorukov~E~V and Loss~D 2000 \PRL {\bf 85} 1962

\end{thebibliography}
\end{document}